\documentstyle[12pt,amsfonts,amssymb,amsbsy,epsf,rotate]{article}
\begin{document}
\makeatletter

\def\section{\@startsection {section}{2}{\z@}{-3.5ex plus -1ex minus
 -.2ex}{2.3ex plus .2ex}{\large\sc}}

\makeatother
\begin{center}
{\large\sf Evolution of flat universe with a cosmological term\\ in modified 
Relativistic Theory of Gravitation\\[1mm] as a scalar-tensor extension of
General
Relativity}
%\end{center}

\vspace*{3mm}
{V.V.Kiselev}\\
\vspace*{3mm}
Russian State Center "Institute for High Energy Physics",\\
Protvino, Moscow region, 142280, Russia\\
E-mail: kiselev@th1.ihep.su, Fax: (0967)-742824\\
Phone: (0967)-713780.
\end{center}

\begin{abstract}
We consider the dynamics of tensor and scalar gravitational fields in the
Relativistic Theory of Gravitation with the Minkowskian vacuum metric and
generalize the formulation to the massless graviton. The potential of scalar
field is determined in the presence of cosmological term under clear physical
motivations. We find cosmological inflationary solutions and analyze conditions
providing the transition to the regime of hot expanding universe.
\end{abstract}

\vspace*{2cm}

\newpage
\section{Introduction}
\begin{figure}[th]
\setlength{\unitlength}{1mm}
%\begin{center}
\hspace*{4cm}{\epsfxsize=8cm \epsfbox{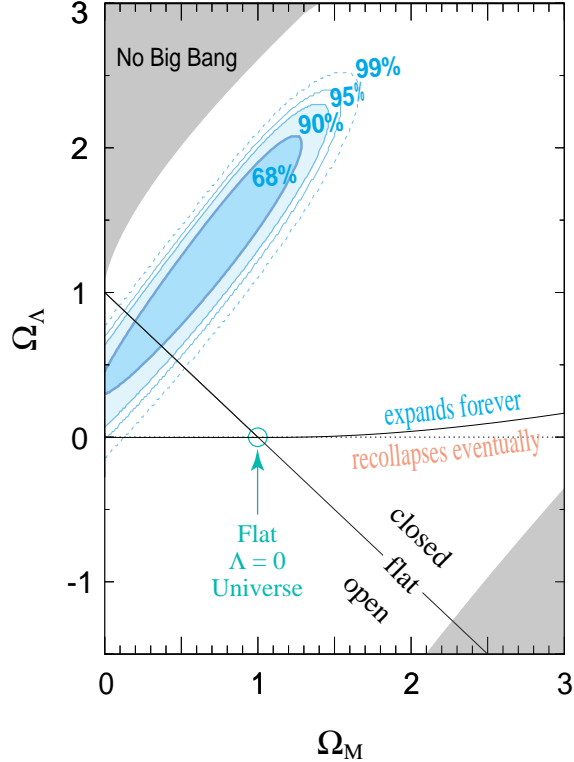}}
%\end{center}
%\caption{}
\caption{\small Results of \cite{Perl} represented in the plane of $\Omega_M$
and $\Omega_\Lambda$ parameters. The flat universe corresponds to the line
$\Omega_\Lambda +\Omega_M=1$.}
\label{fig-omega}
\end{figure}

Recently the astronomical observations on Supernovas with high red shifts
\cite{Perl,Riess} shown that the most probable value for the fraction of
cosmological term in the density of energy differs from zero even in the case
of flat Universe (see Fig. \ref{fig-omega} from \cite{Perl}). Such the
measurements result in the average values of parameters determining the
fractions of energy densities for the matter and cosmological contribution in
ratios to the critical density of energy for the flat Universe, so that
$\Omega_M =0.28^{+0.09}_{-0.08}{}^{+0.05}_{-0.04}$ and $\Omega_\Lambda \approx
0.72$, respectively, with the same error bars.

For the nonrelativistic motion of matter the deceleration parameter of universe
expansion is determined by the expression 
$$
q = \frac{\Omega_M}{2} - \Omega_\Lambda,
$$
which leads to the estimate \cite{Perl,Riess,Perl2}
\begin{equation}
q = -0.33\pm 0.17.
\end{equation}
Thus, the experimental data point to the actuality and necessity of
comprehensive analysis on the evolution of flat universe in the presence of
cosmological term.

The cosmological constant in the flat universe is a necessary ingredient in the
Relativistic Theory of Gravitation (RTG) \cite{RTG}. The construction of RTG by
A.A.Logunov was based on the fundamental physical principle of vacuum
stability, so that, according to RTG, the vacuum is the flat Minkowskian
space-time. In this way the metric tensor of Riemannian space-time is
considered as the Faraday--Maxwell field over the flat vacuum. The field
equations of motion have to satisfy the following principles:

\begin{enumerate}
\item
{\sf The vacuum stability}: the gravitation field equations are identically
true under the absence of matter fields, so that the metric tensor of
Riemannian space-time is equal to the Minkowskian metric.
\item
{\sf The conservation law}: the divergence of energy-momentum tensor for the
gravitational field under the covariant derivative composed by the connection
consistent with the Minkowskian metric is identically equal to zero.
\item
{\sf The weak equivalence principle (the geometrization)}: the interaction of
gravitational field with the matter is built under the Riemannian metric.
\end{enumerate}
The theory of gravitational field constructed in \cite{RTG} leads to the
Lagrangian, which coincides with the Einstein--Hilbert Lagrangian in the theory
of General Relativity, if only the cosmological constant is equal to zero. A
nonzero cosmological term implies the introduction of additional contribution,
which is related with the mass of graviton in RTG. This mass can be introduced
in the presence of background, vacuum metric, the Minkowskian one in the case
under study. In \cite{RTG} the theory with the degenerated masses of tensor
graviton (spin 2) and scalar gravitational field (spin 0) was constructed. The
additional fields of spins 1 and 0 are eliminated from RTG due to the
consequence from the conservation law (the second principle), which makes the
divergence of Riemannian metric under the covariant Minkowskian derivative be
equal to zero. Thus, the RTG is a bi-metric theory reducible to a scalar-tensor
variant for the extension\footnote{Some examples of bi-metric theories of
gravity and scalar-tensor variants are presented in \cite{extended}.} of
standard Lagrangian for the gravitational field in the General Relativity, but
the ambiguity in the choice of functional parameters in the theory is
completely cancelled due to the clear physical restrictions given above.

The motivation of RTG in the light of modern view to the problem is not
restricted by the principle to construct the theory for the Faraday--Maxwell
field of gravitation. Indeed, the geometrization principle for the interaction
of gravity with the matter leads to the nonrenormalizability of quantum theory
as in the General Relativity. However, we can suggest that we deal
with the classical low-energy field theory under virtualities less than the
Planck scale, while a full renormalizable theory is formulated in the flat
space-time\footnote{In the curved space-time the local quantum field theory is
certainly nonrenormalizable.}, and the dimensional Planck scale appears under
the breaking of conformal invariance in the renorma\-lization group and the
spontaneous breaking of higher symmetry in the full theory \cite{Kiselev}. In
this way the stable vacuum in the form of Minkowskian metric can enter the
effective low-energy Lagrangian for the gravitational field, that takes place
in RTG in the case of nonzero cosmological constant. On the other hand, if we
part with the principle of renormalizability, then we can suppose that a full
theory is not local, i.e. it deals with some extended objects, that
necessarily leads to an extension of full high-energy space-time dimension
\cite{M}. These additional dimensions have to be compactified at the energies
below the Planck scale, so that the vacuum is determined by a configuration,
which can enter the effective classic theory of gravity. In the case of flat
four-dimensional vacuum, the Lagrangian can depend on the Minkowskian metric,
that again points to the actuality of consideration with the
theory-construction principles accepted in RTG.

In \cite{RTG} the gravitational field equations are derived in the case of
degenerate masses of tensor and scalar fields. In the study of universe
evolution in that version of RTG, authors found stringent restrictions on the
graviton mass, that followed from the age of Universe. The cosmological
solution in the presence of matter was found pulsating with the
non-inflationary expansion, which has lots of problems \cite{Linde}. Moreover,
the deceleration parameter occurred greater than $\frac{1}{2}$, that
contradicts the mentioned astronomical observations. In general, the rather
strict suggestion on the connection of cosmological constant with the nonzero
masses of gravitational fields can be avoided. In the present paper we modify
the RTG by studying massless graviton and graviscalar in the presence of
nonzero cosmological term. The exploration of RTG principles leads to a
motivated determination of potential for the scalar field, which equation of
motion can be exactly integrated out. In this way the dimensional parameter
giving the cosmological constant is not connected to the masses of
gravitational fields. The offered modification of RTG results in significant
cosmological implications. Namely, the evolution of universe at the initial
stage has the inflationary solution (see a comprehensive description of
inflationary scenario in \cite{Linde}). The value of cosmological constant
falls under the inflation, so that at late times under the transition to flat
expanding universe the deceleration parameter agrees with the recent
astronomical observations. We show at which conditions the equations of motion
in the cosmology lead to the regime of hot expanding universe. 

In Section II we construct the dynamics of modified RTG (MRTG). Then we find
the solutions for the evolution of universe in the presence of cosmological
constant in MRTG with the dust or ultra-relativistic matter in Section III. The
obtained results are summarized and discussed in Conclusion.

\section{Dynamics of tensor and scalar gravitational fields}

According to RTG the density of Riemannian metric ${\mathfrak g}^{\mu\nu} =
\sqrt{-g}\, {g}^{\mu\nu}$ is expressed in terms of sum of densities for the
Minkowskian metric and gravitational Faraday--Maxwell field \cite{Deser} 
\begin{equation}
\begin{array}{rcl}
{\mathfrak g}^{\mu\nu} &=& \sqrt{-\gamma}\, (\gamma^{\mu\nu} + \Phi^{\mu\nu}),
\\[3mm]
D_\mu\, \Phi^{\mu\nu} &=& 0,
\end{array}
\end{equation}
where $D_\mu$ is the covariant derivative with the Christoffel symbols
determined by the Minkow\-skian metric, so that they globally become zero in
the
Galilean (Cartesian) coordinates in the whole space-time, while the zero
divergence actually is the implication of conservation law for the
energy-momentum tensor, that will be shown later. The gravitational field can
be expanded to the traceless tensor part and the scalar contribution, so that 
\begin{equation}
\begin{array}{rcl}
\Phi^{\mu\nu} &=& \left[\Phi^{\mu\nu} - \frac{1}{3}\left(\gamma^{\mu\nu} -
\frac{D^\mu D^\nu}{D^2}\right)\, (\gamma_{\alpha\beta} \Phi^{\alpha\beta}) 
\right] + \frac{1}{3}\left(\gamma^{\mu\nu} - \frac{D^\mu D^\nu}{D^2}\right)\,
\omega^\prime \\[3mm]
&=& u^{\mu\nu} + w^{\prime\mu\nu},\\[3mm]
D_\mu\, u^{\mu\nu} &=& 0,\;\;\;\; \gamma_{\mu\nu}\,u^{\mu\nu} = 0,\\[4mm]
{\mathfrak u}^{\mu\nu} &=& \sqrt{-\gamma}\, u^{\mu\nu} = \sqrt{-\gamma}\,
\left[\Phi^{\mu\nu} - \frac{1}{3}\left(\gamma^{\mu\nu} -
\frac{D^\mu D^\nu}{D^2}\right)\, (\gamma_{\alpha\beta} \Phi^{\alpha\beta}) 
\right],\\[3mm]
{\mathfrak w}^{\prime\mu\nu} &=& \sqrt{-\gamma}\, w^{\prime\mu\nu} =
\sqrt{-\gamma}\,\frac{1}{3}\left(\gamma^{\mu\nu} - \frac{D^\mu
D^\nu}{D^2}\right)\, \omega^\prime.
\end{array}
\end{equation}
We introduce an independent density caused by the scalar field in the form 
\begin{equation}
{\mathfrak w}^{\mu\nu} = {\mathfrak w}^{\prime\mu\nu} + \sqrt{-\gamma}\,
\gamma^{\mu\nu} = \sqrt{-\gamma}\,\frac{1}{3}\left(\gamma^{\mu\nu} -
\frac{D^\mu D^\nu}{D^2}\right)\, \omega,
\end{equation}
where $\omega = 4 + \omega^\prime$, and at zero $\omega^\prime$ we have got the
flat metric\footnote{At the global $\omega$, we have to define the action of
singular operator $D^\mu D^\nu/D^2$. So, contracting $\gamma^{\mu\nu} =
(\gamma^{\mu\nu} - D^\mu D^\nu/D^2)\,\omega_0/3$ with $\gamma_{\mu\nu}$, we
find $\omega_0=4$, that implies the introduction of definition by the
substitution $(D^\mu D^\nu/ D^2)\, \omega \to \gamma^{\mu\nu}\omega/4$ at the
global $\omega$.}. It is significant that the contraction of Riemannian and
Minkowskian metric tensors depends on the scalar component of gravitational
field, only, so that 
$$
\gamma_{\mu\nu} {\mathfrak g}^{\mu\nu} = \gamma_{\mu\nu} {\mathfrak
w}^{\mu\nu},
$$
where, in accordance with our definitions, we put
$$
{\mathfrak g}^{\mu\nu} = {\mathfrak u}^{\mu\nu} + {\mathfrak w}^{\mu\nu}.
$$
The density of Lagrangian for the gravitational field in the modified RTG with
the cosmological constant is expressed in the form\footnote{We accept the
units, in which $4\pi G = 1/m_{\scriptstyle \sc Pl}^2 =1$, $G$ is the
gravitational constant, $m_{\scriptstyle \sc Pl}$ is the Planck mass.}
\begin{equation}
{\mathfrak L} = -\frac{1}{4}\, {\mathfrak R} + \lambda_1 \sqrt{-g} + \lambda_2
\sqrt{-g}\, \gamma_{\mu\nu} w^{\mu\nu} + \sqrt{-\gamma}\,
V(\gamma_{\mu\nu}w^{\prime\mu\nu}),
\end{equation}
where $V(\gamma_{\mu\nu}w^{\prime\mu\nu})$ is an arbitrary function, and we
have not explicitly show the dependence of Riemannian metric on the tensor and
scalar parts. However, we exhibit the Minkowskian metric as well as the
contractions with $w^{\prime\mu\nu}$ and $w^{\mu\nu}$. Of course, we have to
add some terms with the Lagrange multipliers: 
$$
{\mathfrak L}_{\rm mult} = {\mathfrak K}_{\mu\nu}^{[1]}\cdot [a^{\mu\nu}
-\gamma^{\mu\nu}]+ {\mathfrak K}_{\mu\nu}^{[2]}\cdot\left[w^{\prime\mu\nu} -
\frac{1}{3}\left(\gamma_{\mu\nu} - \frac{D_\mu
D_\nu}{D^2}\right)\omega^\prime\right],
$$
where we have put $w^{\mu\nu} = w^{\prime\mu\nu} +a^{\mu\nu}$. The multipliers
allow us to take correct variations of the Lagrangian, while they are not very
essential for the further consideration, since we put the tensor densities
${\mathfrak K}_{\mu\nu}^{[n]}$ equal to zero.

The physical meaning of terms added to the scalar curvature is rather
transparent: we have introduced the variation of energy under the external
source of gravitational field with respect to the flat vacuum. The term with a
factor $\lambda_1$ is the usual cosmological constant in the General
Relativity, and we have to introduce it in accordance with the astronomical
observations. The second term with a factor $\lambda_2$ is chosen in the
minimal model of linear dependence on the contraction of Riemannian and
Minkowskian metrics. We will show that in the minimal model, the ratio
$g/\gamma= \det g_{\mu\nu}/\det \gamma_{\mu\nu}$ will fix the scalar field.
This suggestion on the form of third term significantly restricts the
Lagrangian under study. We will mention on a possible generalization below, but
concentrate the consideration on the minimal model as clarified above.

The minimal term with the factor $\lambda_2$ is necessary in order to satisfy
the first principle of vacuum stability. The term with the potential $V$ is
arbitrary to the moment, but we will present some arguments towards its fixing.

Therefore, the vacuum energy can shift under the source. This phenomenon is
quite ordinary in the field theory, since, for instance, the vacuum energy of
free massive scalar field $\varphi$ is shifted under a source $j$, so that the
potential is given by
$$
U(\varphi, j) = m_\varphi^2\frac{\varphi^2}{2}-j\varphi = 
m_\varphi^2\frac{(\varphi-j/m_\varphi^2)^2}{2}-\frac{j^2}{2m_\varphi^2},
$$
and the density of vacuum energy is misplaced from zero at $j\neq 0$:
$$
U_{\rm vac}(\varphi_{\rm vac}, j) = -\frac{j^2}{2m_\varphi^2},
$$
and it depends on the source.

In MRTG the shift of vacuum energy under the source of Riemannian metric is
introduced in contrast to the General Relativity, where the cosmological term
is constant. The dependence of vacuum energy appears due to the graviscalar
field, and we deal with the scalar-tensor extension of General Relativity.

Then we generally get the equations of motion
\begin{eqnarray}
\frac{\delta{\mathfrak L}}{\delta{\mathfrak u}^{\mu\nu}} & = & 
\frac{\delta{\mathfrak L}}{\delta{\mathfrak g}^{\mu\nu}} -
\frac{1}{3}\left(\gamma_{\mu\nu} -
\frac{D_\mu D_\nu}{D^2}\right)\, \left(\gamma^{\alpha\beta}\,
\frac{\delta{\mathfrak L}}{\delta{\mathfrak g}^{\alpha\beta}}\right),
\label{e1}\\[2mm]
{\displaystyle \frac{\delta{\mathfrak L}}{\delta\omega^\prime}} 
&=&
{\displaystyle \sqrt{-\gamma}\,
\frac{1}{3}\left(\gamma^{\mu\nu} -
\frac{D^\mu D^\nu}{D^2}\right)\, 
\frac{\delta{\mathfrak L}}{\delta{\mathfrak w}^{\prime\mu\nu}}}
\nonumber\\ &=&
{\displaystyle \sqrt{-\gamma}\,
\frac{1}{3}\left(\gamma^{\mu\nu} -
\frac{D^\mu D^\nu}{D^2}\right)\, \left(
\frac{\delta{\mathfrak L}}{\delta{\mathfrak g}^{\mu\nu}}+
\frac{\delta^\star{\mathfrak L}}{\delta{\mathfrak w}^{\prime\mu\nu}}\right),}
\label{e2}
\end{eqnarray}
where the action of $\delta^\star$ implies the variation of terms explicitly
depending on ${\mathfrak w}^{\prime\mu\nu}$ or ${\mathfrak w}^{\mu\nu}$.

In order to match with the General Relativity, we postulate that the form of
field equations for the Riemannian metric is independent of the spin contents
of $g_{\mu\nu}$. This uniform principle of MRTG reads off
\begin{equation}
\frac{\delta{\mathfrak L}}{\delta{\mathfrak g}^{\mu\nu}} = 0.
\label{uniform}
\end{equation}
Then the field equations of motion for ${\mathfrak g}^{\mu\nu}$ have got the
form 
\begin{equation}
\frac{\delta{\mathfrak L}}{\delta{\mathfrak g}^{\mu\nu}} = -\frac{1}{4}\,
R_{\mu\nu} + \frac{1}{2}\, \lambda_1\, g_{\mu\nu} + \frac{1}{2}\, \lambda_2\,
g_{\mu\nu}\, \gamma_{\alpha\beta} w^{\alpha\beta} = 0, 
\end{equation}
According to the RTG principle of vacuum stability, in the limit of
$u^{\mu\nu}\to 0$, $w^{\mu\nu}\to \gamma^{\mu\nu}$ the purely gravitational
equations of motion with no matter have to be identically satisfied, so that we
get 
\begin{equation}
\lambda_1 = -4\lambda_2.
\end{equation}
The graviscalar field equation of motion, taking into account the uniform
principle, gives the expression 
\begin{equation}
\begin{array}{rcl}
{\displaystyle \frac{\delta{\mathfrak L}}{\delta\omega^\prime}} &=&
\lambda_2\, \sqrt{-g} + \sqrt{-\gamma}\,V^\prime(\omega^\prime) = 0, 
\end{array}
\label{motionsc}
\end{equation}
so that in the limit of flat vacuum we find 
\begin{equation}
\left.V^\prime(\omega^\prime)\right|_{\omega^\prime =0} = -\lambda_2.
\end{equation}
Let us emphasize that the uniform principle of equations for the gravitational
field (\ref{uniform}) combined with (\ref{e1}) and (\ref{e2}) implies that the
Riemannian metric $g_{\mu\nu}$ and the graviscalar $\omega$ can be considered
as
independent, since their variations are nor related with each other and enter
in the separate field equations
\begin{eqnarray}
\frac{\delta{\mathfrak L}}{\delta{\mathfrak g}^{\mu\nu}} &=& 0,\\[2mm]
\frac{\delta^\star{\mathfrak L}}{\delta{\mathfrak w}^{\prime\mu\nu}} &=& 0.
\end{eqnarray}
This point is very important, because we have started with the graviscalar
given by the contraction of Riemannian and Minkowskian metrics, but the further
consideration has shown that the dependence of vacuum energy on the external
gravitational source appears due to the connection of graviscalar with the
ratio of Riemannian and Minkowskian metrics as follows from (\ref{motionsc}).
We have found that the modification of RTG leads to the so-called ``slightly
bi-metric'' theory, where the flat metric enters in the full Lagrangian by its
determinant, only.

After taking into account the field equations, the density of energy-momentum
tensor for the gravitational field can be written down in the form
\begin{equation}
\begin{array}{rcl}
{\mathfrak t}^{\mu\nu}_g =-2 {\displaystyle \frac{\delta{\mathfrak
L}}{\delta\gamma_{\mu\nu}} } &=&
{\displaystyle -\frac{1}{4}\, {\mathfrak J}^{\mu\nu} -2\, \lambda_2\,
\sqrt{-g}\, w^{\mu\nu}}\\ && {\displaystyle- \sqrt{-\gamma}\,
V(\gamma_{\alpha\beta}w^{\prime\alpha\beta})\,\gamma^{\mu\nu}-
2\sqrt{-\gamma}\, V^\prime (\gamma_{\alpha\beta}w^{\prime\alpha\beta})\,
w^{\prime\mu\nu}}, 
\end{array}
\end{equation}
where the density of current \cite{RTG} is equal to 
$$
{\mathfrak J}^{\mu\nu} = D_\alpha D_\beta (\gamma^{\alpha\mu}{\mathfrak
g}^{\beta\nu} + \gamma^{\alpha\nu}{\mathfrak g}^{\beta\mu} -
\gamma^{\alpha\beta}{\mathfrak g}^{\mu\nu} -\gamma^{\mu\nu}{\mathfrak
g}^{\alpha\beta}).
$$
In the absence of gravitational field, the vacuum solution has to lead to zero
of introduced energy-momentum tensor, hence, we find 
\begin{equation}
\left.V(\omega^\prime)\right|_{\omega^\prime =0} = -2\lambda_2.
\end{equation}
The graviscalar field equation of motion (\ref{motionsc}) implicates a
stringent restriction to the potential 
\begin{equation}
{\lambda_2}\, \sqrt{-g} = - \sqrt{-\gamma}\, {V^\prime(\omega^\prime)},
\end{equation}
which use provides the following form of energy-momentum tensor:
\begin{equation}
\begin{array}{rcl}
{\mathfrak t}^{\mu\nu}_g &=& {\displaystyle -\frac{1}{4}\, {\mathfrak
J}^{\mu\nu} + \sqrt{-\gamma}\, \gamma^{\mu\nu}\, 
[2V^\prime(\omega^\prime) -V(\omega^\prime)]
.}
\end{array}
\end{equation}
Then the covariant conservation law is valid, if we put
\begin{equation}
\left.\begin{array}{rll} D_\mu\, {\mathfrak g}^{\mu\nu} &=& 0,\\[2mm]
2V^\prime(\omega^\prime) &=& V(\omega^\prime), \end{array} \right\} \;\;\;
\Longrightarrow
\;\; D_\mu\, {\mathfrak t}^{\mu\nu}_g =0.
\end{equation}
The above conditions are sufficient, but necessary. Nevertheless, let us
consider the density of graviton energy-momentum tensor at ${\mathfrak
g}^{\mu\nu} = {\mathfrak w}^{\mu\nu}$, i.e. at zero tensor component. Then the
covariant derivative
\begin{equation}
D_\mu\, {\mathfrak t}^{\mu\nu}_g = \sqrt{-\gamma}\, \gamma^{\mu\nu}\,D_\mu\,
[2V^\prime(\omega^\prime) -V(\omega^\prime)]=0.
\label{truncate}
\end{equation}
For arbitrary $\omega^\prime$ we derive
\begin{equation}
2V^\prime(\omega^\prime) = V(\omega^\prime)
\label{pot-cond}
\end{equation}
under the normalization conditions in the flat space-time. However, the
assumption on zero tensor contribution is quite synthetic, in general, since
it can be inconsistent with the field equation. Therefore, in the modified RTG
the potential of scalar field is determined under the clear motivation, which,
however, does not provide a necessary condition for the conservation law. 

Nevertheless, we can present an additional argument based on the minimal
introduction of flat metric. In that case, one usually expresses the curvature
with no involvement of covariant derivative with the connection consistent with
the Minkowskian metric as ordinary given in the General Relativity. Then the
variation of Lagrangian with respect to the flat metric, i.e. the density of
gravitational energy-momentum tensor, does not include the term of ${\mathfrak
J}^{\mu\nu}$, and eq. (\ref{truncate}) is exact, which implies the strict
derivation of the potential form $V$. Moreover, we emphasize that under both
the General Relativity expression for the curvature and the potential condition 
(\ref{pot-cond}), the energy-momentum tensor of gravitational field is
independent of scalar potential and equal to zero
$$
\left.{\mathfrak t}^{\mu\nu}_g\right|_{\rm GR} = 0,
$$
which is ideologically correct in the purely geometrical theory.

Then in the minimal model of MRTG the motion equation for the scalar field can
be exactly integrated out, so that
\begin{equation}
\begin{array}{rcl}
V(\omega^\prime) &=& -2\lambda_2\,
\exp\left[\frac{1}{2}\omega^\prime\right],\\[3mm]
\omega^\prime &=&\displaystyle \ln \left[\frac{g}{\gamma}\right].
\end{array}
\end{equation}
After the substitution, the transversity of tensor density ${\mathfrak
t}^{\mu\nu}_g$ becomes explicit 
\begin{equation}
\frac{1}{4}\, D^2 {\mathfrak g}^{\mu\nu}  = {\mathfrak t}^{\mu\nu}_g,
\end{equation}
if
$$
D_\mu\, {\mathfrak g}^{\mu\nu} =0.
$$
whereas for the trace we obtain the expression
\begin{equation}
D^2\,(\gamma_{\mu\nu}\, {\mathfrak g}^{\mu\nu}) = {4}\,
{\mathfrak t}^{\mu\nu}_g\, \gamma_{\mu\nu}.
\end{equation}
At small perturbations of gravitational field both the tensor and scalar
components are massless, and the dimensional parameter determining the
cosmological constant is not related to a nonzero mass of gravitons. So, we
denote
$$
\lambda_1 = \frac{m_\Lambda^2}{2},\quad
\lambda_2 = - \frac{m_\Lambda^2}{8},
$$
while the masses are equal to zero:
$$
m_u^2 = 0, \quad m_\omega^2 =0.
$$
This result could be expected, since the field equation of graviscalar allows
us to express the determinant of Minkowskian metric in terms of both the
determinant of Riemannian metric and the scalar field. This fact implies that
in MRTG one can make the Minkowskian metric to enter the Lagrangian implicitly,
and the only appearing of $\gamma$ is the determination of graviscalar value as
well as the condition of transversity. This fact provides the breaking of
strong equivalence principle by the inherent dependence on the value of
graviscalar field related with the Minkowskian metric. We explain that this
dependence is due to the shift of vacuum energy because of the external
gravitational field, and this shift is the only point of such the breaking,
while the weak equivalence of gravity and inertia is valid.

Let us make a note on the generalization of consideration by the substitution
$$
\lambda_2 \sqrt{-g}\, \gamma_{\mu\nu}w^{\mu\nu}\to 
\lambda_2 \sqrt{-g}\, \tilde V(\gamma_{\mu\nu}w^{\mu\nu}),
$$
where $\tilde V(4)=4$ for the definiteness. In that case, the form of relation
between $\lambda_1$ and $\lambda_2$ holds with no change, as well as the
derivation concerning for the potential $V$ and the massless gravitational
fields remains valid. The field equation of graviscalar reads off
\begin{equation}
\lambda_2\sqrt{-g}\, \tilde V^\prime(\gamma_{\mu\nu}w^{\mu\nu})+
\sqrt{-\gamma}\,V^\prime(\gamma_{\mu\nu}w^{\prime\mu\nu})=0.
\label{ext}
\end{equation}
Therefore, if the potential $\tilde V$ is arbitrary, the relation between
the graviscalar and the ratio of metric determinants is indefinite. Moreover,
some choices of $\tilde V$ could be inconsistent with the existence of
meaningful solution of (\ref{ext}). Nevertheless, we can argue for the linear
dependence of $\tilde V$. Indeed, the quantum loop corrections to the
gravitational Lagrangian due to the matter fields result in the scale
dependence of cosmological constant (see, for instance, \cite{birreldavis}).
So, the scalar field with a mass $m$ gives the following contribution:
\begin{equation}
\left.m_{\Lambda}^2\right|_{\rm eff} = \left.m_{\Lambda}^2\right|_{\rm bare} -
\frac{m^4}{4\pi^2 m_{\scriptstyle\sc Pl}^2}\, \ln\frac{\mu_0}{\mu},
\end{equation}
The scale factor of ${\mu_0}/{\mu}$ can be evidently replaced by the
ratio of curved and flat metrics:
$$
\ln\frac{\mu_0}{\mu} = \frac{1}{8}\,\ln\frac{g}{\gamma} + {\rm const.}
$$
where we have explored a simple relation between the differentials 
$$
{\rm d}\ln \mu = - {\rm d}\ln \Omega,
$$
where $\Omega$ is a dilution factor resulting in the scaling of metrics
$g_{\mu\nu} = \Omega^2\,\gamma_{\mu\nu}$ in the case of conformal relation. The
arbitrary constant in the above relation of scale factors can be fixed by the
principle of vacuum stability, since we require that the vacuum is flat, and
hence,
$$
\left.m_{\Lambda}^2\right|_{\rm eff} = -\frac{m^4}{32\pi^2 m_{\scriptstyle\sc
Pl}^2}\, \ln\frac{g}{\gamma}.
$$
Thus, the cosmological term should tend to zero in the vacuum state, while its
dependence on the scale factor expressed in terms of the determinants is given
by the renormalization procedure, i.e. the corresponding function is the log.
Of course, this dependence can be modified by higher orders, a summation of
perturbative contributions by various matter fields, which can change the sign
of factor in front of the logarithm. Nevertheless, we stress
that in the leading order one could use the logarithmic variation of the
cosmological term, that fixes the form of $\tilde V$ as the linear function.
This fact reminds of the minimal prescription in the theory under study.

We stress that the above procedure giving the dependence of the cosmological
term on the external gravitational source can be involved in the modified RTG,
but the General Relativity, since this action breaks the strong equivalence
principle in what concerns for the vacuum, which can enter the low-energy
effective action.

\begin{figure}[th]
\setlength{\unitlength}{1mm}
\begin{center}
\begin{picture}(100,60)
\put(3,2){\epsfxsize=8cm \epsfbox{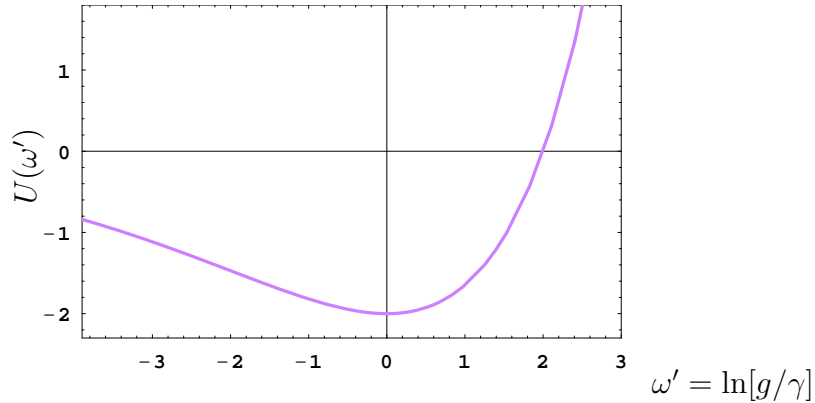}}
\put(0,25){\rotate{$U(\omega^\prime)$}}
\put(85,0){$\omega^\prime=\ln [{g}/{\gamma}]$}
\end{picture}
\end{center}
%\caption{}
\caption{\small The dependence of scalar field potential in dimensionless
units as follows from (\ref{dimenu}).}
\label{fig-u}
\end{figure}

Then the density of Lagrangian for the gravitational field can be represented
in the form
\begin{equation}
\begin{array}{rcl}
{\mathfrak L} &=& \displaystyle -\frac{1}{4}\, {\mathfrak R} + \sqrt{-\gamma}\,
[4 V^\prime(\omega^\prime) - V^\prime(\omega^\prime)\,\omega +V(\omega^\prime)]
=\\[4mm]
&=&\displaystyle  -\frac{1}{4}\, {\mathfrak R} + \frac{m_\Lambda^2}{8}
\sqrt{-\gamma}\, (2 - \omega^\prime) \exp\left[\frac{1}{2}\omega^\prime\right].
\end{array}
\end{equation}
The effective potential of scalar field in the flat Minkowskian space-time is
equal to
\begin{equation}
V_{\rm eff} = - \frac{m_\Lambda^2}{8}
\sqrt{-\gamma}\, (2 - \omega^\prime) \exp\left[\frac{1}{2}\omega^\prime\right]
=  \frac{m_\Lambda^2}{8}\, \sqrt{-\gamma}\, U(\omega^\prime),
\label{dimenu}
\end{equation}
where the dimensionless function $U(\omega^\prime)$ has got the only minimum
corresponding to the flat vacuum of Riemannian space-time (see Fig.
\ref{fig-u}).

In terms of Riemannian metric, the potential can be transformed to the form
\begin{equation}
V_{\rm eff} = - \frac{m_\Lambda^2}{8}
\sqrt{-g}\, \left(2 - \ln \left[\frac{g}{\gamma}\right]\right),
\end{equation}
where the vacuum Minkowskian metric enters explicitly.

It is evident that the geometrization principle for the interaction of
gravitation with the matter provides the validity of above conclusions on the
form of gravitational Lagrangian. The motion equations are written down as
\begin{equation}
R_{\mu\nu} - \frac{1}{2}\, g_{\mu\nu} R - \frac{m_\Lambda^2}{4} \ln
\left[\frac{g}{\gamma}\right]\,g_{\mu\nu} = 2\, T^M_{\mu\nu},
\label{basic}
\end{equation}
where the energy-momentum tensor of matter is defined under its Lagrangian
density ${\mathfrak L}^M$ by the expression
$$
T^M_{\mu\nu} = 2\,\frac{\delta{\mathfrak L}^M}{\sqrt{-g}\delta
g^{\mu\nu}}.
$$
Thus, we have completely determined the dynamics of gravitational field in RTG
with the cosmological constant.

Finally, we give the formula for the Lagrangian density of gravitational field
in the modified RTG
\begin{equation}
{\mathfrak L}_{\rm MRTG} = -\frac{1}{4}{\mathfrak R} - \frac{m_\Lambda^2}{8}\,
\sqrt{-g}\, \omega^\prime + {\mathfrak K}\cdot \left[\omega^\prime - \ln
\frac{g}{\gamma}\right],
\label{MRTG}
\end{equation}
where the external graviscalar field is fixed after the variation over the
Lagrange multiplier $\mathfrak K$. The action with (\ref{MRTG}) results in the
field equations (\ref{basic}) under the summation with the action of matter. In
addition, if one follows the ideology of General Relativity with the minimal
involvement of flat vacuum metric, the harmonic condition
$$
D_\mu\, {\mathfrak g}^{\mu\nu} = 0,
$$
is treated as the gauge condition fixing the arbitrary connection between the
flat and Riemannian metrics, while in MRTG (nonzero ${\mathfrak J}^{\mu\nu}$)
this is the consequence of conservation law (the second principle).

Two comments are to the point. First, we emphasize the different role of mass
parameter in the modified and original versions of RTG. Certainly, in the
theory under consideration the tensor graviton field is massless as it is in
the General Relativity, while in the Logunov's RTG it is massive as well as the
dynamical scalar field. In the modified RTG, the dimensional parameter
$m_\Lambda$ can have a very different dynamical nature, since it is connected
to the contribution of gravitational degrees of freedom into the
energy-momentum tensor. So, in cosmological models the scale parameter can
depend on both properties of physical vacuum in a full theory and interactions
of gravitons with the matter. Another example could be considered in problems
on a gravitational field with a point-like or rotating sources, where the scale
parameter can depend on the position with respect to the source point or its
axis. Moreover, the position dependence of $m_\Lambda$ could be adjusted to
give some reasonable limits for such problems. Thus, we have the different
treatment of scale parameter in the original RTG and its modified version under
study. We stress ones more that the scale parameter in the modified RTG is
related with the cosmological constant, however, it is not straightforwardly
associated with the graviton or graviscalar mass, since the tensor graviton is
massless, while the scalar field equation is algebraic, and hence, it is not
dynamical.

The second comment concerns for the bimetric feature. Following authors of
\cite{PittsSchieve}, we emphasize that the so-called ``slightly bimetric''
theory involving the determinant of Minkowskian metric into the Lagrangian
only, is equivalent to the scalar-tensor gravity. Those bimetric theories in
\cite{PittsSchieve} are constructed under the same or very close principles
accepted in RTG. So, the universal coupling of gravity, conservation law and
presence of Minkowskian metric are declared. Then, an additional free parameter
of Lagrangian is a flat cosmological term with the Minkowskian metric. The
gauge principle is explored to show that the energy-momentum tensor is fixed if
only the gauge condition is chosen. In the slightly bimetric theories the gauge
invariance is restricted, indeed. Such constructions are equivalent to general
covariant theories plus a scalar field, so that every slightly bimetric theory
has a scalar-tensor ``twin'' and {\it vice versa}. Thus, we do not see any
constructive problem in the formulation of such bimetric theory. Another
problem is an interpretation of bimetric theories since we deal with two null
cones associated with the propagation of interactions. Indeed, for the particle
physicists one considers the graviton as the interaction carrier with respect
to the flat vacuum, i.e. the Minkowskian metric. Therefore, the principle of
causality states that the propagation should be inside the null cone of this
flat metric, which implies that the null cone of curved Riemannian metric
should be inside the timelike region of Minkowskian one. However, this
causality principle proceeds the inequality condition depending on the gauge
change of Riemannian metric. One could to restrict the gauge arbitrariness to
get some conclusions. Nevertheless, the problem reveals to be more complex. It
is connected to the general interpretation for the case of intersecting null
cones. The first opinion is supported by A.A.Logunov, who declared that all of
solutions with the intersection of null cones are not physical, since only the
causal propagation is permitted. The objections to that point of view are based
on both the restricted gauge invariance arbitrary changing the sign in the
causality inequality and the presentation of special solutions, which are
physically reasonable, but demonstrate the intersections of null cones. For
example, solutions for the gravitational field of massive rotating bodies far
from the source (the so-called Lense-Tirring metrics) with the angular momentum
large enough result in that the inclination (in the asimutal direction) of null
cone for the Riemannian metric with respect to the null cone of Minkowskian
metric can become so large even at large distances from the body that these
cones intersect each other. Another opinion insists that the flat metric is a
fiction because of the null cone problems. Both versions of interpretation
suffer from the gauge-noninvariant formulation of arguments. Related questions
are extensively discussed in \cite{PittsSchieve} provided by a complete list
of references. We can add the third point of view: the problem of causality in
terms of null cones could be fictitious. Indeed, what one can observe is the
null cones for the motion in the Riemannian metric, while the Minkowskian null
cones are the almost-vacuum limits of Riemannian ones, so that the only
quantity accessible for the observation in the slightly bimetric theory as MRTG
(with the ratio of determinants for the flat and curved metrics) is the
compression or dilution factor for the Riemannian space-time with respect to
the Minkowskian one, if we suggest the gauge invariance of this factor. Anyway,
in what follows, we deal with the Friedmann--Robertson--Walker metric, which is
conformally flat, and therefore, the characteristic surfaces, i.e. the gravity
and matter null cones coincide for this particular case. We will show the
importance of cosmological term in this situation, while in other applications
the value of scale parameter, in fact, can be put to zero due to some classical
motivations like the minimization of energy or absence of hot gravitational
field and so on.

Finally, let us make a note on the covariant conservation of energy-momentum
tensor. In the General Relativity one has
\begin{equation}
\nabla_\mu g^{\mu\lambda}T^M_{\lambda\nu} =0.
\label{divRTG}
\end{equation}
The physical meaning of (\ref{divRTG}) is the following: at arbitrary external
gravitational source $g_{\mu\nu}$ interacting with the matter, the invariance
of matter action with respect to the general coordinate transformations under
the field equations leads to the conservation law. The introduction of vacuum
energy, i.e. the cosmological constant, in the General relativity results in
the conservation of 
$$
\tilde T^M_{\lambda\nu} = T^M_{\lambda\nu}+ \Lambda g_{\lambda\nu}.
$$
However, under the connection consistent with the Riemannian metric
$$
\nabla_\mu g_{\lambda\nu} = 0,
$$
eq. (\ref{divRTG}) remains valid since the energy-momentum tensors of both the
matter and the vacuum are conserved separately.

In MRTG the full tensor $\hat T^M_{\mu\nu}$ of matter and vacuum is conserved,
so that
\begin{equation}
\nabla_\mu g^{\mu\lambda}\hat T^M_{\lambda\nu}=\nabla_\mu
g^{\mu\lambda}\left(T^M_{\lambda\nu}+ 
\frac{m_\Lambda^2}{8} \ln
\left[\frac{g}{\gamma}\right]\,g_{\lambda\nu}\right)=0.
\label{divMRTG}
\end{equation}
The physical meaning of (\ref{divMRTG}) is certainly clear: under the
introduction of external source in the form of gravitational field, the density
of vacuum energy changes, i.e. the value of cosmological term depends on the
Riemannian metric. Therefore, considering the action of matter in the external
gravitational field, one has to take into account the vacuum contribution
depending on the Riemannian metric, so that the variation of full action for
the matter and the vacuum under the general coordinate transformations is equal
to zero if (\ref{divMRTG}). 

\section{Cosmological solutions}
In this section we analyze the scenario for the universe evolution in MRTG.
From the very beginning it is clear that in MRTG due to the principle of vacuum
stability the empty flat space-time is the solution of motion equations, and
this fact will be illustrated in this section, so that the expansion of
universe containing a matter takes place under an introduction of some
instability. We will show that such the instability appears if the energy
densities of the matter and cosmological term are approximately equal to each
other, and the universe is compressed with respect to the flat vacuum. Then the
expansion takes place if we consider the Lagrangian density with the negative
square of mass parameter introduced above. Thus, the starting point of our
analysis is the gravitation Lagrangian density written down in the form 
\begin{equation}
{\mathfrak L} = -\frac{1}{4}\, {\mathfrak R} + \frac{\bar\mu_\Lambda^2}{8}
\sqrt{-g}\, \omega^\prime + {\mathfrak K}\cdot \left[\omega^\prime - \ln
\frac{g}{\gamma}\right].
\end{equation}
Consider the Riemannian metric, which gives the following
Friedmann--Robertson--Walker interval:
\begin{equation}
ds^2 = dt^2 - a^2(t)\, [dr^2 + r^2 (d\vartheta^2 + \sin^2\vartheta
d\varphi^2)].
\label{FRW}
\end{equation}
The proper-time presentation of Riemannian metric in (\ref{FRW}) is consistent
with the field equations of RTG ({\it viz.}, $D_\mu {\mathfrak g}^{\mu\nu}=0$)
\cite{RTG}, if the Minkowskian metric has the form
\begin{equation}
ds_\gamma^2 = \frac{1}{a^6(t)}\, dt^2 - \frac{1}{\varkappa^2} [dr^2 + r^2
(d\vartheta^2 + \sin^2\vartheta d\varphi^2)],
\label{flatFRW}
\end{equation}
where $\varkappa$ is a global scale factor.

The cosmological term can be rewritten as the contribution into the
energy-momentum tensor 
\begin{equation}
T^\Lambda_{\mu\nu} = 2\,\frac{\delta}{\sqrt{-g}\delta g^{\mu\nu}}\,\left\{
\frac{\bar\mu_\Lambda^2}{8} \sqrt{-g}\, \, \omega^\prime
\right\}_{\omega^\prime = \ln \frac{g}{\gamma}} = -
\frac{\bar\mu_\Lambda^2}{8} \ln \left[\frac{g}{\gamma}\right]\,g_{\mu\nu}.
\end{equation}
The equations of motion (\ref{basic}) with the metric under study are
transformed to the form 
\begin{equation}
\begin{array}{rcl}
\displaystyle \frac{\ddot a}{a} &=& \displaystyle  -\frac{1}{3}\, (\rho + 3
p),\\[3mm]
H^2 &=& \displaystyle \frac{2}{3}\, \rho,
\end{array}
\label{dot}
\end{equation}
where we have introduced the standard notation for the Hubble constant $H =
\dot a/a$ and written the energy-momentum tensor in terms of energy density
$\rho$ and pressure $p$. The covariant conservation of energy-momentum tensor
leads to the liquid equation, which follows from (\ref{dot})
\begin{equation}
\dot\rho a^3 +3(\rho+p) a^2\dot a =0.
\end{equation}
For the given metrics (\ref{FRW}) and (\ref{flatFRW}) we can easily find that
$\ln g/\gamma = 12\ln a\sqrt{\varkappa}$. However, without a matter we expect
that
the cosmological constant is equal to zero, and we approach the stable vacuum
solution, that implies $a^2(t)\, \varkappa \equiv 1$. Moreover, adjusting the
limit
of free vacuum $g_{\mu\nu}\to \gamma_{\mu\nu}$, we get the condition $\varkappa
= 1$.
Further, we introduce the notation $\mu^2_{\Lambda} = 2 \bar\mu^2_{\Lambda}$,
which corresponds to the naive scale behaviour of $\ln g \sim 6 \ln a$.

Let us study the case of dust matter with the cosmological term in MRTG. Then
\begin{eqnarray}
\rho_\Lambda &=& -\frac{3}{4}\, \mu_\Lambda^2\,\ln a,\;\;\; p_\Lambda =
-\rho_\Lambda,\\ 
\rho_M &=& \frac{3}{4}\, \mu_\Lambda^2\,\rho_0 ,\hspace*{9mm} p_M = 0,
\end{eqnarray}
where we have introduced the dimensionless parameter $\rho_0$.  Equations
(\ref{dot}) lead to the expressions
\begin{eqnarray}
\dot H &=& -\rho_M,\\
H^2 &=& \frac{\mu_\Lambda^2}{2}\, \left[\ln \frac{1}{a} +
\rho_0\right],
\end{eqnarray}
and
\begin{equation}
\dot\rho_0 =  H\, (1-3\rho_0).
\end{equation}
Putting $\rho_0$ equal to its stable value\footnote{Here it is important that
$H>0$.} at late times we find that the matter density is constant, and 
\begin{equation}
\rho_0=\frac{1}{3},\;\;\;\; \rho_M = \frac{\mu_\Lambda^2}{4}.
\end{equation}
Then the time-equations are easily integrated out, so that
\begin{equation}
\begin{array}{rcl}
a(t) &=& \displaystyle \frac{1}{k}\, \exp\left[-\left(\sqrt{-\ln(a_0
k)}-\frac{{\mu_\Lambda}}{2\sqrt{2}} t\right)^2\right],\\[5mm]
H^2 &=& \displaystyle \frac{\mu_\Lambda^2}{2}\,\left(\sqrt{-\ln(a_0
k)}-\frac{{\mu_\Lambda}}{2\sqrt{2}} t\right)^2,\\[5mm]
\dot H &=& \displaystyle -\frac{\mu_\Lambda^2}{4},
\end{array}
\end{equation}
where $k= \exp[-\rho_0]$, and the initial data at zero time have to satisfy the
condition $\ln(a_0 k)<0$.

We see that the universe has got the exponential, inflationary expansion with
the constant density of matter, which is significantly less than the density of
cosmological term at early times of initial stage. The cosmological term with
respect to the flat Minkowskian space-time is naturally interpreted as the
contribution of hot strongly coupled gravitons. The density of gravitons falls
with the time of inflation, while the entropy is transmitted from the gravitons
to the matter, so that the matter entropy grows exponentially.

The deceleration parameter $q = - \frac{\ddot a}{a}\frac{1}{H^2}$ is determined
by the relation
\begin{equation}
q = -1 - \frac{\dot H}{H^2},
\end{equation}
and at the initial stage it is close to $-1$, while in the end of inflation it
becomes close to the value observed experimentally. So, for example, we get 
$$
\left.q\right|_{(\rho_0-\ln a) =1} = - \frac{1}{2},\;\;\;
\left.q\right|_{(\rho_0-\ln a) =3/4} = - \frac{1}{3}.
$$
The initial conditions of inflation for the flat universe in MRTG are
determined at the Planck scale, so that we can naturally suggest that
fluctuations are in the following range:
$$
\begin{array}{rcl}
\delta a &\sim & a,\;\;\; a\to 0,\\[2mm]
\delta t &\sim & \displaystyle\frac{1}{m_{\scriptscriptstyle \sc Pl}},
\end{array}
$$
hence,
$$
H^2_0 \sim m_{\scriptscriptstyle \sc Pl}^2 \;\Longrightarrow \;\;
\ln\frac{1}{a_0}\sim \frac{m_{\scriptscriptstyle \sc Pl}^2}{\mu_\Lambda^2}.
$$
This fact implies that at low values of dimensional parameter in MRTG the
inflation has got a huge scale, that is necessary for an achievement of
homogeneous and isotropic thermal equilibrium before the transition to the
regime of hot expanding universe (see \cite{Linde}).

For the relativistic matter
\begin{equation}
\rho_{\scriptscriptstyle RM} = \frac{3}{4}\,
\mu_\Lambda^2\,\rho_{\scriptscriptstyle 0R} ,\hspace*{9mm}
p_{\scriptscriptstyle
RM} = \frac{1}{3}\,\rho_{\scriptscriptstyle RM},
\end{equation}
we can write down analogous expressions, such as, for example,
\begin{equation}
\dot\rho_{\scriptscriptstyle 0R} =  H\, (1-4\rho_{\scriptscriptstyle 0R}).
\label{relro}
\end{equation}
Choosing the stable value, we get 
\begin{equation}
\rho_{\scriptscriptstyle 0R}=\frac{1}{4},\;\;\;\; \dot H =
-\frac{\mu_\Lambda^2}{4}= -\frac{4}{3}\,\rho_{\scriptscriptstyle RM} .
\end{equation}
Other expressions for the solution remain valid after the corresponding
redefinition of normalization parameter for the relativistic case, so that
$k_{\scriptscriptstyle R} = \exp[-\rho_{\scriptscriptstyle 0R}]$.

Let us stress that the given solutions for the cosmological evolution have been
obtained with no reference to the statistical thermal properties of gravitons
and matter. For example, let us consider the relations, which take place for
the relativistic gas of non-interacting particles. Their densities of energy
$\rho _M$ and entropy $s_M$ versus the temperature $T$ are given by the
expressions
\begin{equation}
\begin{array}{rcl}
\rho_M &=& \displaystyle \frac{\pi^2}{30}\, {\cal N}\, T^4,\\[3mm]
s_M &=& \displaystyle \frac{2\pi^2}{45}\, {\cal N}\, T^3,
\end{array}
\label{rel-state}
\end{equation}
where the number of degrees of freedom is equal to the sum over the
polarization states of bosons and fermions with the masses less than the
temperature, so that 
$$
{\cal N} = {\cal N}_B+\frac{7}{8}\, {\cal N}_F.
$$
The presented cosmological solutions show that the ultra-relativistic ideal gas
has the constant density of energy, and, hence, it has got a constant
temperature. Thus, the inflationary expansion takes place isothermically for
the matter, therefore its total entropy increases as $S_M = s_M\,a^3(t)\,V_0$.
It is evident that this process can occur only under the loss of entropy by
gravitons. The qualitative picture of universe inflation is the following: in
the starting point of inflation at the Planck scale or so, the gravitons have
got the large density in comparison with the dilute matter, so that the strong
gravitational self-interaction of gravitons leads to a small scattering length
(probably about the Planck length), and the gravitons interact with each other,
while the matter density is small, and the transparent matter, in practice,
evolves isothermically, since the scattering length of gravitons in the matter
is much greater than the self-interaction length of gravitons. Therefore, in
this approximation we can suppose that the matter, ``in practice,  weakly''
interacts with the gravitational field. In the end of inflation the density o
gravitons decreases to the values close to the parameters of matter, i.e. the
universe little deviates from the flat one with the Minkowskian metric.
Actually, the graviton scattering length off graviton becomes approximately
equal to the scattering length off matter. At this point, the matter state
equation differs from that of relativistic particles weakly interacting with
the gravity. Probably, the warm gravitons heat up the matter, until the
graviton density becomes too low to make a little influence on the cosmological
evolution. We see that the universe inflation results in the dominant
interaction of gravitons with the matter. Indeed, the scattering length is
defined in terms of cross section $\sigma$ and particle density $n$
$$
\lambda = \frac{1}{\sigma\,n},
$$
therefore, if we naively estimate the particle density by the energy density
divided by the temperature
$$
n \approx n_0\,\frac{\rho}{T},
$$
and put the cross section equal to
$$
\sigma = \left.{\rm const}\, \frac{\alpha_{\sc\scriptscriptstyle
GR}^2}{T^2}\right|_{\alpha_{\sc\scriptscriptstyle
GR}=\frac{T^2}{m_{\sc\scriptscriptstyle Pl}^2}} ={\rm const}\,
\frac{T^2}{m_{\sc\scriptscriptstyle Pl}^4},
$$
then we find the ratio of graviton scattering lengths off the matter and
gravitons
$$
\frac{\lambda_M}{\lambda_{\sc\scriptscriptstyle GR}} =
\frac{T_{\sc\scriptscriptstyle GR}(t)\, \ln \frac{1}{a(t)}}{T_M\, \frac{1}{4}},
$$
where the dependence of graviton temperature on the time
$T_{\sc\scriptscriptstyle GR}(t)$ is determined by the state equation:
density-temperature-volume, for the gravitons. So, we see that, if the
temperature of gravitons is close to that of matter, then the tending of scale
factor $a$ to unity during the inflation leads, initially, to the equalizing of
scattering lengths, and further to the dominant interaction of gravitons with
the matter.

If the total entropy of gravitational field and matter is conserved, then we
have got the following expression for the entropy per a unit volume $V_0$ (in
comoving coordinates independent of time, see the definition of interval):
$$
S_0 = a^3(t)\, s_M(T)+S_{\sc\scriptscriptstyle GR}.
$$
Further, we write down the standard relation for the total energy of
gravitational field
\begin{equation}
d{\cal E}_{\sc\scriptscriptstyle GR} = T_{\sc\scriptscriptstyle GR}\,
dS_{\sc\scriptscriptstyle GR} - p_{\Lambda}\, dV, 
\label{therm}
\end{equation}
where, evidently,
\begin{equation}
\begin{array}{rcl}
V(t) &=& \displaystyle a^3(t),\\[2mm]
{\cal E}_{\sc\scriptscriptstyle GR} &=& \displaystyle \rho_\Lambda (t)\,
a^3(t),\\[2mm]
p_\Lambda &=& - \rho_\Lambda,
\end{array}
\label{vol}
\end{equation}
so that after taking the derivative with respect to time we find
$$
T_{\sc\scriptscriptstyle GR} = - \frac{V(t)\, \dot\rho_\Lambda}{s_M(T)\, \dot
V(t)} = \frac{\mu_\Lambda^2\,m_{\sc\scriptscriptstyle Pl}^2}{4s_M(T)}.
$$
Therefore, we have got that the temperature of gravitons also does not change
during the time of inflation
$$
T_{\sc\scriptscriptstyle GR} = \frac{45}{8\pi^2 {\cal N}}\,
\frac{\mu_\Lambda^2\,m_{\sc\scriptscriptstyle Pl}^2}{T^3},
$$
and the process is isothermic for the gravitons, too. Moreover, comparing the
state equation of matter (\ref{rel-state}) with the stable density of matter in
the inflation 
$$
\rho_M = \frac{3}{16}\, \mu_\Lambda^2\, m_{\sc\scriptscriptstyle Pl}^2,
$$
we find that 
$$
T_{\sc\scriptscriptstyle GR} = T,
$$
i.e. the temperature of gravitons is equal to the temperature of matter during
the inflation.

The equality of temperatures and the adiabaticy of process can be obtained
also from the general equation of state (\ref{therm}) combined for the
gravitons and matter, if we take into account (\ref{vol}) and analogous
equations for the relativistic matter with the stable density. Indeed,
calculating the differentials in the equation
$$
d{\cal E}_{\sc\scriptscriptstyle GR}+d{\cal E}_M = (T_{\sc\scriptscriptstyle
GR}-T_M)\, dS_{\sc\scriptscriptstyle GR} - p_{\Lambda}\, dV- p_{M}\, dV, 
$$
we get 
$$
(T_{\sc\scriptscriptstyle GR}-T_M)\, dS_{\sc\scriptscriptstyle GR} = 0,
$$
so that, if the process is adiabatic for the closed system, then the
temperatures of gravitons and matter are equal to each other.

Further, we have to emphasize two circumstances. First, at $H<0$, i.e. after
the end of inflationary isothermic expansion and the transition to the
contraction, the point of constant density of matter
$\rho_{\sc\scriptscriptstyle 0R}$ becomes {\bf unstable} in accordance with
(\ref{relro}). Second, for the gauge non-gravitational interactions of matter
with the coupling constant $\alpha$ we can estimate the cross section by 
$$
\sigma_{\rm gauge} = \sigma_0\, \frac{\alpha^2}{T^2},
$$
so that the scattering length in such interactions of matter is significantly
less than the graviton scattering length off matter,
\begin{equation}
\lambda_{\rm gauge} \sim \frac{1}{\alpha^2\, T}\ll \lambda_M
\sim \frac{m_{\sc\scriptscriptstyle Pl}^4}{T^5},
\label{leng}
\end{equation}
This fact implies that in the beginning of inflation, when the graviton
scattering off gravitons dominates, the isothermic adiabatic expansion takes
place, while in the end of inflation, when the graviton scattering length off
matter closes to the self-scattering length, the gravitons begin to lose the
energy and heat up the matter, so that the absorbed energy of gravitons is
transformed into the thermal motion due to the gauge interactions because of
(\ref{leng}). Therefore, both the density and entropy of matter grow in the
comoving coordinates. This increase is permitted, since the stability of
constant value for the matter density is destroyed. By the process of such
heating, at the time $t_1$ the energy density takes the form 
\begin{eqnarray}
\rho_{\sc\scriptscriptstyle I} &=& \frac{3}{4}\,
\tilde\mu_{\sc\scriptscriptstyle I}^2\,m_{\sc\scriptscriptstyle
Pl}^2\, \left(\ln\frac{1}{a_1\,k_{\sc\scriptscriptstyle R}} +\frac{\Delta
\rho_{\sc\scriptscriptstyle IR}}{a_1^4} \right),\\
\rho_{\sc\scriptscriptstyle II} &=& \frac{3}{4}\,
\tilde\mu_{\sc\scriptscriptstyle II}^2\,m_{\sc\scriptscriptstyle
Pl}^2\, \left(\ln{a_1\,k_{\sc\scriptscriptstyle R}} +\frac{\Delta
\rho_{\sc\scriptscriptstyle IIR}}{a_1^4} \right),
\end{eqnarray}
where $a_1=a(t_1)$, and by (\ref{relro}) we put
\begin{eqnarray}
\rho_{\sc\scriptscriptstyle IR}(t) &=& \frac{1}{4} + \frac{\Delta
\rho_{\sc\scriptscriptstyle IR}}{a^4(t)},\;\;\;\; 
\rho_{\sc\scriptscriptstyle IM} = \frac{3}{4}\,
\tilde\mu_{\sc\scriptscriptstyle I}^2\,m_{\sc\scriptscriptstyle
Pl}^2\, \left(\frac{1}{4} +\frac{\Delta \rho_{\sc\scriptscriptstyle IR}}{a^4}
\right),\\
\rho_{\sc\scriptscriptstyle IIR}(t) &=& \frac{1}{4} - \frac{\Delta
\rho_{\sc\scriptscriptstyle IIR}}{a^4(t)},\;\;\;\; 
\rho_{\sc\scriptscriptstyle IIM} = \frac{3}{4}\,
\tilde\mu_{\sc\scriptscriptstyle II}^2\,m_{\sc\scriptscriptstyle
Pl}^2\, \left(-\frac{1}{4} +\frac{\Delta \rho_{\sc\scriptscriptstyle IIR}}{a^4}
\right),
\end{eqnarray}
where $\Delta \rho_{\sc\scriptscriptstyle R}$ is the constant of integration,
which we define in accordance with two variants of process development, since
the parameter $\tilde \mu^2$ modified due to the graviton absorption can take
two signs: in the second variant this parameter changes the sign in comparison
with its state before the absorption $\mu_\Lambda^2$, while in the first
variant
the sign remains with no change. The physical sense of these two variants will
be shown below under the description of their cosmological differences.

The nature of $\mu^2$ fall off may be twofold. First, we stress the essential
fact of instability for the matter density, so that the absorption of gravitons
can play a role of stabilization, and under the change of sign
$\mu_\Lambda^2\to
-\tilde\mu_{\sc\scriptscriptstyle II}^2$ the stability of potential is
restored. Second, it may be important that the changing scale of evolution
$a(t)$ corresponds to an introduction of renormalization group scale dependence
of matter charges and Yukawa-like couplings, so that the physical phase state
rearrangement is possible, since such critical parameters as the temperature of
phase transition, in general, depend on the mentioned running coupling
constants.

A powerful absorption of gravitons leads to 
$$
\left|\ln\frac{1}{a_1\,k_{\sc\scriptscriptstyle 0R}}\right| \ll \frac{\Delta
\rho_{\sc\scriptscriptstyle R}}{a_1^4},
$$
and we get the evolution equation
\begin{equation}
H^2 = \frac{1}{2}\, \tilde\mu_{\sc\scriptscriptstyle
I,II}^2\,m_{\sc\scriptscriptstyle Pl}^2\,
\frac{\Delta \rho_{\sc\scriptscriptstyle I,IIR}}{a^4(t)},
\end{equation}
that implies the transition to the scenario of hot expanding universe, because
\begin{equation}
a^2(t) = a_1^2 +\sqrt{2}\, \tilde\mu_{\sc\scriptscriptstyle
I,II}\,m_{\sc\scriptscriptstyle Pl}\, (t-t_1),
\end{equation}
and at late times for the relativistic matter we obtain 
$$
a(t)\sim \sqrt{t}.
$$
The temperature of hot universe $T_{hot}$ is determined from the comparison of
energy density in the thermal equilibrium with the expression given in terms of
$\rho_{\sc\scriptscriptstyle 0R}$, so that 
$$
\frac{\pi^2}{30}\, {\cal N}\, T^4_{hot} \approx \frac{3}{4}\,
\tilde\mu_{\sc\scriptscriptstyle I,II}^2\,m_{\sc\scriptscriptstyle Pl}^2\,
\frac{\Delta \rho_{\sc\scriptscriptstyle R}}{a_1^4} ,
$$
wherefrom we have got an ordinary relation
$$
T(t) \sim \frac{1}{a(t)}.
$$
Since the absorption of gravitons by the matter takes place at the constant
volume and with no loss of energy, we have introduce the notation $\tilde\mu$
for the dimensional parameter after the absorption, so that, evidently,
comparing the total energy before the absorption of gravitons and after it, we
get
$$
\tilde\mu_{\sc\scriptscriptstyle I}^2 \approx \tilde\mu_{\sc\scriptscriptstyle
II}^2 \approx \mu_{\Lambda}^2\,
\frac{\ln\frac{1}{a_1\,k_{\sc\scriptscriptstyle
R}}}{\frac{\Delta \rho_{\sc\scriptscriptstyle R}}{a_1^4}}\;\; \Rightarrow\;\;
\tilde\mu^2 \ll \tilde\mu_\Lambda^2.
$$
Thus, the stage of hot expanding universe can be quite long in time in order
to have no contradiction with the astronomical observations.

The hierarchy of scales $\mu_\Lambda$ and $\tilde\mu$ implies also that at the
graviton absorption 
$$
\mu_\Lambda^2\, \frac{1}{4} \approx \tilde\mu^2\,
\frac{\Delta\rho_{\sc\scriptscriptstyle R}}{a_1^4},
$$
and the heating up the matter is little in comparison with the stage of
inflation.

Further, the given consideration can be analogously performed for the dust
matter having a zero pressure. Then we get 
$$
\rho_{0}(t) = \frac{1}{3} \pm \frac{\Delta
\rho_{0}}{a^3(t)},
$$
and the law of hot universe expansion is ordinary again
$$
a(t) \sim \root 3\of{t^2}.
$$

Let us also consider the problem on the reaching the ``critical'' zero density
of matter in the second scenario of postinflationary evolution. Indeed, at
$a(t)> \root 4\of {4\Delta\rho_{\sc\scriptscriptstyle IIR}}$ the density of
matter formally becomes negative, that has no physical sense for nonzero modes,
if we do not suppose that this situation is possible under, first, the vacuum
energy is negative in the state equations of matter, and, hence, the null-point
of energy measure is posed at a negative energy; second, it is necessary
that $\ln a \gg 1$ and a correct cosmological solution could take place: $H^2
=\frac{2}{3} \rho>0$, though, as clear from the astronomical observations, 
this condition is valid.

The notion on the connection of matter density with the nonzero value of
effective potential in the state of matter vacuum is important, because we see
that at the inflation the stable isothermic density of matter energy
$V_M^{inf}(0) \sim \mu^2_\Lambda m_{\sc\scriptscriptstyle Pl}^2>0$, while in
the
hot universe the matter density tends to a stable value of
$V_{\sc\scriptscriptstyle IM}(0) > 0$ or $V_{\sc\scriptscriptstyle IIM}(0)<0$
depending on the variant of development.

Finally, we note also that the one-loop quantum corrections lead to a
renormalization of cosmological constant \cite{birreldavis}, so that in MRTG
they determine the adiabatic dependence of dimensional parameter $\mu_\Lambda$
on the scale $a$, since it can be related with the ultraviolet cut off $M$ by
the formula $d\ln M = -d\ln a$. In this work we do not discuss a motivation and
numerical estimates of dimensional parameters for the cosmological evolution in
MRTG.

A short remark should be done on a value of matter density fluctuations, which
determine the large scale structure of Universe. It is clear that, if the
graviton scattering length off matter is close to the scattering length in the
gauge interactions of matter, then the fluctuations are about $\delta\rho/\rho
\sim 1$. At the moment of inflation transition to the regime of hot universe we
expect that 
$$
\frac{\delta\rho}{\rho} \approx \frac{\lambda_{\rm gauge}}{\lambda_M}\sim
\frac{1}{\alpha^2}\,\left(\frac{c_{\cal N} T}{m_{\sc\scriptscriptstyle
Pl}}\right)^4.
$$
The astronomical observations agree with the altitude of initial fluctuations
$$
\frac{\delta\rho}{\rho} \sim 10^{-(4-5)},
$$
that leads to 
$$
\frac{T}{m_{\sc\scriptscriptstyle Pl}} \sim 10^{-(2-4)},
$$
if $\alpha\sim 10^{-2}$ and the numerical factor $c_{\cal N}$ in front of
temperature can be about ${\cal N}\sim 1-10^{2}$.

Thus, we have analyzed the cosmological solutions in the modified RTG and found
the conditions of inflationary expansion as well as transition to the ordinary
evolution of hot universe.

\section{Conclusion}

In the present paper we have shown that the cosmological scenario of universe
evolution possesses some advantages in the modified Relativistic Theory of
Gravitation, which naturally introduces the dependence of gravitational field
action on the vacuum metric of Minkowski at nonzero cosmological term necessary
in accordance with the astronomical observations. Certainly, the evolution
equations allow the development of logically sound stages providing the
inflationary expansion and the modern epoch of late hot Universe. In this way
the inflation takes place isothermically, and we have not to introduce a
mechanism for a forced heating up a overcooled homogeneous and isotropic
universe under the transition from the inflation to the hot universe. In MRTG
the following stages of cosmological evolution are theoretically sound:

1. The beginning: a hot state with a Planck-scale density of both
ultra-relativistic matter and gravitons; $\rho_M\approx \rho_\Lambda\sim
m_{\sc\scriptscriptstyle Pl}^4$ with a temperature of
$T_{\star\sc\scriptscriptstyle GR} \sim T_{\star M} \sim
m_{\sc\scriptscriptstyle Pl} \sim 10^{19}$ GeV in a strongly compressed state 
$\ln\frac{1}{a_\star}\gg 1$; a time $t_\star$.

2. The preinflation: a stage of expansion and matter cooling down a temperature
$T_M$, so that $T_M^4 \sim \mu_\Lambda^2 m_{\sc\scriptscriptstyle Pl}^2$, where
$\mu_\Lambda^2\sim \frac{m_{\sc\scriptscriptstyle Pl}^2}{-\ln a_\star}\ll
m_{\sc\scriptscriptstyle Pl}^2$; a scale of $T_M\sim 10^{14-16}$ GeV; the time
changes from $t_\star$ to $t_0$, whereas $\ln\frac{1}{a(t_0)}\gg 1$.

%\vspace*{1.7mm}
3. The isothermic hot inflation with $T_{\sc\scriptscriptstyle GR}=T_M$ up to a
time $t_1$, when the density of gravitons falls off a level of matter density 
$|\ln a(t_1)|\sim 1$.

\begin{figure}[th]
\setlength{\unitlength}{1mm}
\begin{center}
\begin{picture}(100,100)
\put(-30,0){\epsfxsize=16cm \epsfbox{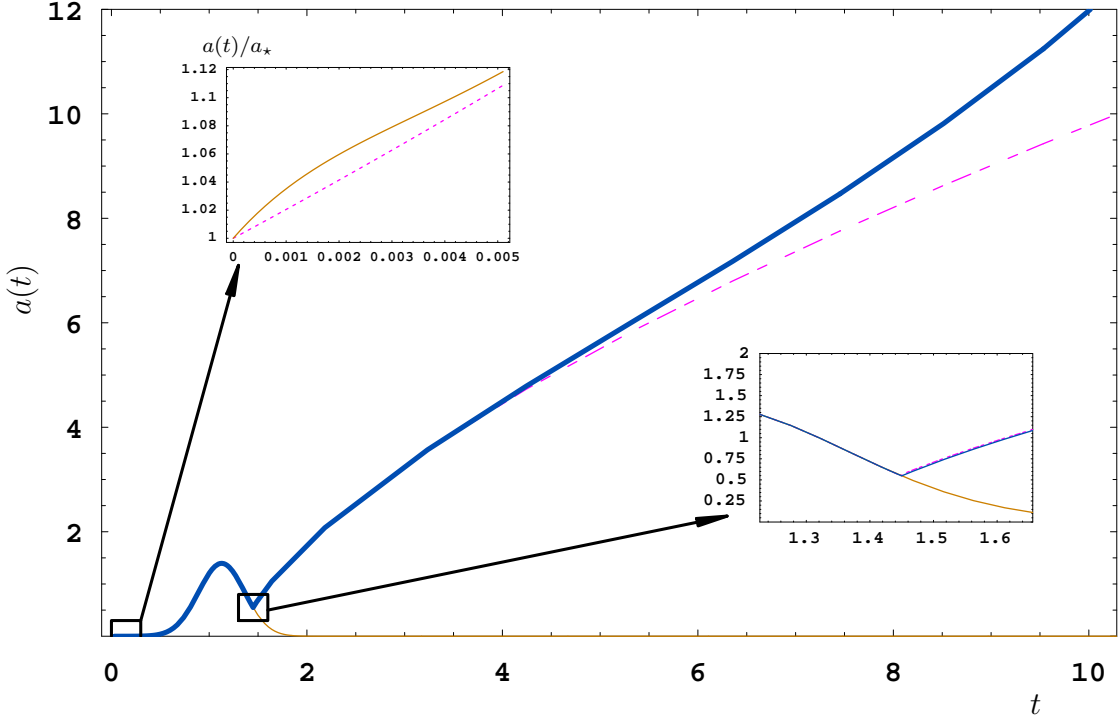}}
\put(-26,53){\rotate{$a(t)$}}
\put(0,88){$\scriptstyle a(t)/a_\star$}
\put(110,0){$t$}
\end{picture}
\end{center}
%\caption{}
\caption{\small Stages of universe evolution in accordance with items 1-6. In
the begin stage we show the scenario of cooling down below the Planck
temperatures before the inflation in comparison with the pure inflation (the
dotted line). The unstable regime change takes place under the graviton
absorption to the process of hot universe expansion in comparison with the
rapid exponential contraction. At late times we present the deviation from the
regime of hot universe expansion (the dashed curve) due to a cold inflationary
expansion (the solid curve). The time $t$ and scale $a(t)$ units are modelled
and far away from the real ones.}
\label{figevol}
\end{figure}

4. The short stage of instability $H<0$, when the energy of gravitons is
converted to the thermal energy of matter due to the absorption of gravitons,
so that before the absorption $\rho_{\sc\scriptscriptstyle GR}(t_1) \sim
\rho_M(t_1)$, while after it $\tilde\rho_{\sc\scriptscriptstyle GR} \ll
\tilde\rho_M$; the restoration of stability\footnote{For the relativistic
matter the evolution equations: $$H^2 = \frac{2}{3}(\rho_M +
\rho_{\sc\scriptscriptstyle GR}),\;\; \frac{\ddot a}{a} =
H^2-\frac{4}{3}\rho_M,$$ result in the inflection point at $\rho_{\sc
\scriptscriptstyle GR}=\rho_M,$ while after the graviton absorption we get
$\tilde\rho_M\gg \tilde\rho_{\sc\scriptscriptstyle GR},$ so that the transition
from the concave regime of contraction to the convex curve of
expansion is possible (see Fig. \ref{figevol}).} $H>0$, whereas the matter is
slightly heated up, and the evolution parameter $\tilde\mu^2/\mu_\Lambda^2 \sim
\tilde\rho_{\sc\scriptscriptstyle GR}/\rho_{\sc\scriptscriptstyle GR}\ll
1$.

5. The present: the epoch of hot universe expansion with the matter
temperature $T(t)\sim\frac{1}{a(t)}$ down to a temperature $T_c$, so that
$T_c^4 \sim\tilde\mu^2 m_{\sc\scriptscriptstyle Pl}^2$. Today we are ``close''
to the end of this stage because in the case of nonrelativistic cosmological
motion of matter the deceleration parameter 
$$
q = -1 + \frac{3}{2}\frac{\rho_M}{\rho_M+\rho_{\sc\scriptscriptstyle
GR}}\approx -0.33\pm 0.17,
$$
gives the witness that $\rho_{\sc\scriptscriptstyle GR} \sim (1-2)
\rho_M$, i.e. the matter density is critically low.

6. The future II: a stable density of matter energy $V_M(0)<0$, the infinite
cold isothermic inflation with a logarithmically small density of matter. The
speed of inflation here could be ``damped'' by a parametric adiabatic decrease
of parameter $\tilde\mu^2$ down to zero because of a renormalization
scale-dependent behaviour, that suggests the tending of matter vacuum density
to zero, $V_M(0)\to 0$. This scenario of future with $V_M(0)<0$ is the most
probable in the light of astronomical data on the deceleration parameter, which
is negative at the moment.

The future I: a stable density of matter energy $V_M(0)>0$, the expansion of
universe up to the equalizing of matter density with the negative contribution
of cosmological term, which provides $q>\frac{3}{2}$, that contradicts the
current data, the later stage of instability of cold universe $H<0$, so that
whether the restoration of stability will lead to an expansion again as in
items
4-5, but for the overcooled universe with a further repetition of transition
to a future according to item 6, or the contraction of universe will take place
with significant fluctuations of density for the matter heated up.

Thus, in MRTG we have got quite the whole picture of universe evolution.

\vspace*{3mm}
The author expresses his gratitude to prof. M.A.Mestvirishvili for fruitful
discussions, prof. V.A.Petrov for an encouraging interest to the studies and
thanks Dr. V.O.Soloviev for questions, remarks and suggestions.

This work is in part supported by the Russian Foundation for Basic Research,
grants 01-02-99315, 01-02-16585 and 00-15-96645.

\end{document}